\documentclass[aps,twocolumn,floatfix,showpacs]{revtex4} 
\usepackage{amsmath,amsfonts,amssymb,graphicx,times,bbm,natbib}
\usepackage[T1]{fontenc}
\usepackage{times}
\newcommand{\id}{\mathbbm{1}}
\DeclareMathOperator{\tr}{tr}
\DeclareMathOperator{\circM}{circ}

\newcommand{\trnorm}[1]{\lVert#1\rVert_\text{tr}}
\newcommand{\abs}[1]{\lvert#1\rvert}
\newcommand{\vect}[1]{\mathbf{#1}}

\begin{document}

\title{Entropy, entanglement, and area: \\%
analytical results for harmonic lattice systems}

\author{M.\ B.\ Plenio}

\affiliation{QOLS, Blackett Laboratory, Imperial College London,
Prince Consort Road, London SW7 2BW, UK}

\author{J.\ Eisert, J.\ Drei{\ss}ig, and M.\ Cramer}%

\affiliation{Institut f{\"u}r Physik, Universit{\"a}t Potsdam,
Am Neuen Palais 10, D-14469 Potsdam, Germany}


\begin{abstract}
We revisit the question of the relation between entanglement,
entropy, and area for harmonic lattice Hamiltonians  corresponding
to discrete versions of real free Klein-Gordon fields. For the
ground state of the $d$-dimensional cubic harmonic lattice we
establish a strict relationship between the surface area of a
distinguished hypercube and the degree of entanglement between the
hypercube and the rest of the lattice  analytically,
without resorting to numerical means. We outline extensions of
these results to longer ranged interactions, finite temperatures
and for classical correlations in classical harmonic lattice
systems. These findings further suggest that the tools of quantum
information science may help in establishing results in quantum
field theory that were previously less accessible.
\end{abstract}

\date{\today }
\maketitle

Imagine a distinguished geometrical region of a discretized free
quantum Klein-Gordon field: what is the entropy associated with a
pure state obtained by tracing over the field variables outside
the region?  How does this entropy  relate to properties of the
region, such as volume and boundary area? This innocent-looking
question is a long-standing  issue indeed, studied in the
literature under the key word of geometric entropy. Analytical
steps supplemented by numerical computations for half-spaces and
spherical configurations in seminal works by Bombelli  \textit{et
al.}\ \cite{Bombelli KLS 86} and Srednicki \cite{Srednicki 93}
strongly suggested  a direct  connection between entropy and area.
The interest in this quantity for quantum field theory is drawn
from the fact that geometric entropy is thought to be the leading
quantum correction to the Bekenstein-Hawking black hole entropy
\cite{BlackHoleStuff}. Subsequent work employed various
approaches, such as methods from conformal field theory
\cite{Holzhey 95}, analysis of entropy subadditivity \cite{Casini
04} or mode counting \cite{Yurtsever 03}.  Recently, there has
been renewed interest in studying entanglement and correlations in
quantum many-body systems and quantum field theory, largely due to
availability of novel powerful methods from the quantitative
theory of entanglement in the context of quantum information
theory \cite{Stelmachovic B 01,Summers W 85,Audenaert EPW
02,Fannes HM 03,Wolf VC 04,Botero R 04,Latorre}. Such ideas have
previously been employed to assess the entanglement in settings of
one-dimensional spin (see, e.g.,  Refs.~\cite{Fannes HM
03,Latorre})  and harmonic chains \cite{Audenaert EPW
02,Botero R 04}.

This letter gives an analytical answer to the question of the
scaling of the degree of entanglement for harmonic lattice
Hamiltonians  such as discrete versions of the free scalar
Klein-Gordon field, in arbitrary spatial dimensions. Although we
encounter a highly correlated system we nevertheless find an
`area-dependence' of the degree of entanglement.  Our analysis is
based on methods that have been developed in recent years in
quantum information theory, in particular those relating to
entanglement in Gaussian (quasi-free) states (see, e.g.,
Ref.~\cite{Eisert P 03}). These methods allow us to give an
analytical answer to the question of the scaling of the degree of
entanglement between a region and its exterior for harmonic
lattice Hamiltonians such as discrete versions of the free scalar
Klein-Gordon field, in arbitrary spatial dimensions. It is
remarkable that although we encounter a highly correlated system,
we nevertheless find an ''area dependence'' of the degree of
entanglement.

{\it The Hamiltonian. ---} The starting point of the argument is a
discrete lattice version of a free real scalar quantum field. For
any $d\ge 1$ we consider a $d$-dimensional simple cubic lattice
$n^{\times d}$ comprising $n^d$ oscillators.
We may write the Hamiltonian as
\begin{equation}\label{hamiltonian}
    H= \vect{p}\vect{p}^\text{T} /2+\vect{x}V \vect{x}^\text{T}/2,
\end{equation}
where $\vect{x}=(x_1,\dots,x_{n^d})$ and
$\vect{p}=(p_1,\dots,p_{n^d})$ denote the canonical coordinates of
the system. The $n^d\times n^d$-matrix $V$, the potential matrix,
specifies the coupling between the oscillators in the position
coordinates.
\begin{figure}[th]\vspace*{0.1cm}
\centerline{
\includegraphics[width=5.5cm]{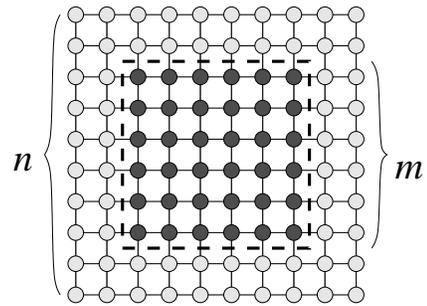}
}
\vspace{.0cm} \caption{\label{fig}  The harmonic lattice in $d=2$
with a distinguished $m\times m$ region in an $n\times n$
lattice.}\vspace*{0.15cm}
\end{figure}

For now $V$~will be chosen such that in the continuum limit one
obtains the Hamiltonian of the real Klein-Gordon field, under
periodic boundary conditions. We will therefore consider the
harmonic lattice Hamiltonian with nearest-neighbor interaction.
Note that our argument can be extended to other types of
interactions. The case of next-to-nearest-neighbor coupling will
also be  discussed later in this paper,  see
Ref.~\cite{Long} for a more general discussion.

We write $ V=\circM(\vect{v})$ for the circulant matrix whose
first row is given by the $n$-tupel $\vect{v}$, and also for a
block circulant matrix where the first block column is specified
by a tupel of matrices. So in $d=1$, we have $V_{1}=\circM(1,
-c,0,\dots,0,-c)$  and in higher dimensions we have a
recursive, block-circulant structure reflecting rows, layers etc.:
\begin{equation*}
    V_{d}=\circM( V_{d-1}, -c
    \id_{n^{d-1}},0,\dots,0,  -c \id_{n^{d-1}})
\end{equation*}
with a $0\le 2cd<1$.  From now on we will write $V$ instead of
$V_d$.

{\it Entanglement and area dependence. ---} We denote the ground state of
the system by $\rho$. For a distinguished cubic
region  $m^{\times d}$ in a lattice $n^{\times d}$ (see Fig.~\ref{fig}) its
entropy of entanglement is
\begin{equation*}
    E_{n,m}=-\tr  \rho_{n,m} \log \rho_{n,m}.
\end{equation*}
The reduced density matrix $\rho_{n,m}$ is formed by tracing out
the variables outside the region $m^{\times d}$. We will
show the following:
 \smallskip

\noindent  \emph{ The entropy of
entanglement of the distinguished region $m^{\times d}$   in
the lattice  $n^{\times d}$ satisfies
\begin{equation}\label{claim}
    \lim_{n\rightarrow\infty}
    E_{n,m} =\Theta( m^{d-1}),
\end{equation}
where $\Theta$ is the Landau-theta. More specifically,
we have that
$
   C_{1}\, m^{d-1} \leq E_{n,m} \leq C_{2}\, m^{d-1}$
   for sufficiently large $m$,
   with appropriate $C_{1},C_{2} > 0$.}

The `area dependence' manifests itself as follows: For a linear
chain, the entropy of entanglement is bounded by quantities that
are independent of the size of the distinguished interval. In two
dimensions, this dependence is linear in the length of the
boundary, in three dimensions to the area of the boundary. Indeed,
one can show that while  all oscillators are correlated with all
oscillators, the correlations over the boundary decay very
quickly. In effect, for fixed interaction strength \footnote{Note
that the continuum limit requires a careful analysis as the
ensuing limit $c\rightarrow 1/2d$  is concomitant to a diverging
correlation length. For a one-dimensional set there is strong
evidence for a logarithmic dependence of the entropy of
entanglement in the continuum limit \cite{Botero R
04,Latorre,Long}.}, the only significant contribution comes from
within a finite width, the correlation length, along the boundary,
and thus leads to a  surface dependence of the
correlations. This intuition forms the basis of the following,
fully analytical argument, where the above statement is proven by
finding upper and lower bounds for which the statement holds.

\smallskip
{\it The upper bound. ---} The ground state $\rho$ of  the coupled
harmonic system in  Eq.~\eqref{hamiltonian} is a Gaussian
(quasi-free) state with vanishing first moments. The second
moments of $\rho$ can be collected in the covariance matrix
$\gamma$, which is defined as $\gamma_{j,k}= 2\text{Re}[ R_j R_k
\rho ]$   for $j,k=1,\dots,2 n^d$, where
$\vect{R}=(x_1,\dots,x_{n^d}, p_1,\dots,p_{n^d})$ is the vector of
canonical coordinates. In terms of the potential matrix $V$ the
covariance matrix  of the ground state  is then found to be
$\gamma= V^{-1/2} \oplus V^{1/2}$  \cite{Audenaert EPW 02}. From
entanglement theory we know that an upper bound for the entropy of
entanglement is provided by the  logarithmic negativity $E_N =
\ln\trnorm{\rho^\Gamma}$,  where $\rho^\Gamma$ is the partial
transpose of  $\rho$, and $\trnorm{\cdot}$ denotes the trace norm
\cite{Neg}. Following Ref.~\cite{Audenaert EPW 02}  we find
\begin{equation}\label{LogNeg}
    E_N= \sum_{j=1}^{n^d}
    \ln(1+\max(0,\lambda_j(Q-\id))),
\end{equation}
where $\lambda_{j}(Q)$ are the non-increasingly ordered
eigenvalues of the matrix
\begin{equation}\label{q}
    Q=V^{-1/2} P V^{1/2} P.
\end{equation}
In a reordered list of canonical coordinates (such that the inner
oscillators are counted first) $P$ is the diagonal matrix $P=-\id_{m^d}\otimes\id_{n^d-m^d}$ and the potential matrices can be written  as
\begin{eqnarray*}
    V^{-1/2} =
    \begin{bmatrix}
    A & B \\
    B^\text{T} &C
    \end{bmatrix},\,
    V^{1/2} =
    \begin{bmatrix}
    D & E \\
    E^\text{T} & F
    \end{bmatrix},\,
    T = \begin{bmatrix}
        0 & E \\
        E^\text{T} & 0
    \end{bmatrix}.
\end{eqnarray*}
The matrices $B$ and $E$ describe the couplings between the
$m^d$ oscillators forming the distinguished hypercube and the rest
of the lattice. On using $V^{-1/2} V^{1/2}=\id$, we arrive at
\begin{eqnarray*}
    Q-\id = -2 V^{-1/2} T
\end{eqnarray*}
This is convenient as it will turn out that the detailed structure
of $V^{-1/2}$ will not have to be considered and we can
concentrate on the properties of the matrix $T$. To avoid taking
the maximum in  Eq.~\eqref{LogNeg} we bound the eigenvalues
by their absolute values,
\begin{align*}
    E_N &\leq \sum_{j=1}^{n^d}\ln(1+|\lambda_j(Q-\id)|)\\
    &\leq\sum_{j=1}^{n^d}|\lambda_j(Q-\id)|=\trnorm{Q-\id},
\end{align*}
where we  have employed that $\ln(1+x)\leq x$ for all $x\ge 0$.
Since the trace norm is unitarily invariant   \cite{Matrix},
we may further write
\begin{equation*}
\trnorm{Q-\id}=2\ \trnorm{V^{-1/2} T} \leq 2\ \lambda_{1}(V^{-1/2})\ \trnorm{T}.
\end{equation*}
Here we also have that $V^{-1/2}$ is symmetric. The spectrum of $V$ can be
obtained via discrete Fourier transform and yields
$\lambda_{1}(V^{-1/2} )=(1-2cd)^{-1/2}$.

Now the trace norm of $T$ can be bounded from above by the sum of the
absolute values of all the matrix elements of $T$, which is known
as the $l_1$ matrix norm \cite{Matrix}. Therefore,
\begin{eqnarray*}
    E_N\leq \frac{2}{\sqrt{1-2cd}}\sum_{i,j=1}^{n^d} \abs{T_{ij}}.
 \end{eqnarray*}
In the following we will bound the matrix elements of $V^{1/2}$
and consequently those of $T$. The explicit implementation of the
multidimensional discrete Fourier transform is non-technical yet
involved. To achieve a more compact notation, 
we introduce the lattice coordinate vectors $\vect{k}, \vect{l}$
where $k_{j},l_{j}=0,\dots,n-1$ and $j=0,\dots,d-1$. For the
considered lattice structure we may write $V_{\vect{k},\vect{l}}=
V_{\sum_{j=0}^{d-1} k_j n^j,\sum_{j=0}^{d-1} l_j n^j}$ for the
interaction term between site $\vect{k}$ and $\vect{l}$.  The
matrix elements of $V^{1/2}$ are then given by
\begin{eqnarray*}
    V^{1/2}_{\vect{k},\vect{l}}
    &=& \sum_{\vect{k}'} \prod_{j=0}^{d-1}
    \frac{e^{ 2\pi i k'_j(k_{j}-l_j)/n}}{n^d}
    \biggl(1-2c\sum_{r=0}^{d-1} \cos \frac{2\pi
    k_r'}{n}\biggr)^{1/2}.
\end{eqnarray*}
To bound these, we replace the square root by its power series
expansion in the parameter $2c$. This converges if $2cd\le 1$,
which coincides with the constraint imposed by the positivity of
the potential matrix. We will use $(1-x)^{1/2} =
-\sum_{s=1}^{\infty} B_s x^s$, with $0<B_s<1$ and the fact that
$\sum_{q=1}^{n} e^{{2\pi i pq}/{n}} = 0$ for integer $p$ and $q$
unless $p$ is a multiple of $n$.  With this the non-diagonal
elements of $V^{1/2}$, and analogously $V^{-1/2}$, are bounded by
\begin{align}
    \frac{y^{ s(\vect{k},\vect{l})}}{1-y } \ge V^{-1/2}_{\vect{k},\vect{l}} \ge 0 \geq V^{1/2}_{\vect{k},\vect{l}} \ge
    -&\frac{y^{ s(\vect{k},\vect{l})}}{1-y},\label{v+}
\end{align}
where
$s(\vect{k},\vect{l})=(k_{0}-l_{0})+\dots+(k_{n-1}-l_{n-1})$,
$y=2cd$ and $0\leq  k_j- l_j\leq n/2$. This demonstrates the
exponential decay of the off-diagonal elements in these block
circulant matrices. The remaining matrix elements are determined
by the periodic boundary conditions under the exchange $k_j- l_j
\mapsto n- (k_j - l_j)$.  Note that generally
$s(\vect{k},\vect{l})$ is simply the number of lattice steps one
has to make starting at site $\vect{k}$ to reach site $\vect{l}$.
If, for example, $s(\vect{k},\vect{l})=1$ then the oscillators are
direct neighbors.

We may now proceed with the computation of the $l_1$ norm of $T$,
i.e., of the blocks in $V^{1/2}$ that describe the coupling
between the distinguished region and the rest of the lattice.
Given that the region is a hypercube, this can be done in a
transparent way.   Consider the set ${\cal L}_0$ of
$m^d-(m-2)^d$ oscillators of the hypercube that lie directly on
the boundary and successively the sets ${\cal L}_r$ of
$(m-2r)^d-(m-2r-2)^d$ oscillators inside that are exactly $r$
steps away from the surface of the hypercube. Starting from the
set ${\cal L}_0$ and taking $s$ steps on the lattice one can reach
less than $(m+2s)^d-m^d$ oscillators outside the hypercube
$m^{\times d}$. Therefore we find that the sum of all the elements
of $T$ that couple oscillators from the set ${\cal L}_0$ to
oscillators outside the hypercube is bounded by
\begin{equation*}
    S_0 \le 2\sum_{s=1}^{\infty} ((m+2s)^d-m^d) \frac{y^s}{1-y} \, .
\end{equation*}
Now consider the contribution from the set ${\cal L}_k$. Clearly,
any oscillator outside the hypercube that can be reached from
${\cal L}_k$ in $s+k$ steps can be reached from ${\cal L}_0$ in
$s$ steps. Therefore we can bound the sum $S_k$ of all the
elements of $T$ that couple the set ${\cal L}_k$ to oscillators
outside the hypercube by
\begin{eqnarray*}
    S_k &\le& 2\sum_{s=k+1}^{\infty} ((m+2(s-k))^d-m^d)
    \frac{y^s}{1-y}.
\end{eqnarray*}
As a consequence we obtain
\begin{eqnarray*}
    E_N &\le& \frac{2}{\sqrt{1-2cd}} \sum_{k=0}^{m/2} S_k\\
    &\le& \frac{2}{\sqrt{1-2cd}} \sum_{s=1}^{\infty} ((m+2s)^d-m^d)
    \frac{y^s}{1-y}\sum_{k=0}^{m/2} y^k.
\end{eqnarray*}
Using the binomial expansion of $(m+2s)^d$ and the Gamma-function
to bound expressions of the form $\sum_{s=0}^{\infty}y^s(2s)^k$ we
find for $m>4d/|\ln(y)|$ the bound
\begin{eqnarray}
    E_N &\le& \frac{16 d }{\sqrt{1-2cd}(1-2cd)^2|\ln(1-2cd)|^2}\, m^{d-1},
\end{eqnarray}
which is the desired upper bound that is linear in the number of
oscillators on the surface of the hypercube.

{\it Lower bound. ---} In the following we demonstrate that the
degree of entanglement, measured by the entropy of entanglement,
is asymptotically  at least linear in the number of
oscillators. The entropy of entanglement depends only on the
symplectic spectrum of the covariance matrix $\gamma_A$
corresponding to the reduced Gaussian state of the interior. The
non-increasingly ordered symplectic eigenvalues satisfy
$\mu_i=(\lambda_i(AD))^{1/2}\ge 1$ from which the entropy of
entanglement can be evaluated as
\begin{eqnarray*}
S &=& \sum_{i=1}^{m^d}\left( \frac{\mu_i+1}{2} \log
\frac{\mu_i+1}{2} - \frac{\mu_i-1}{2} \log \frac{\mu_i-1}{2}
\right).
\end{eqnarray*}
For $\mu_i>1$ each bracketed terms in the sum can be bounded from
below by $\log\mu_i$. Because  $\mu_i\le\left(
(1+2c)/(1-2c)\right)^{1/4}$  for all $i$, we find
\begin{equation*}
 S\ge \sum_{i=1}^{m^d}\log \left( 1+\left(\mu_i-1\right)\right)
 \ge \frac{\log \mu_1}{\mu_1-1}\sum_{i=1}^{m^d} \left(\mu_i-1\right)\, .
\end{equation*}
Employing that for $\beta=(\sqrt{1+k}-1)/k$ we have $\sqrt{1+x}\ge
1+ \beta x$ in $x\in [0,k]$ and $\lambda_i(AD)=1+\lambda_i(-BE^\text{T})$
we find
\begin{equation*}
    S \ge \frac{\sqrt{1+\lambda_1(-BE^\text{T})}-1}{\lambda_1\left(-BE^\text{T}\right)}
    \frac{\log \mu_1}{\mu_1-1} \tr\left(-BE^\text{T}\right).
\end{equation*}
The factors in front of the trace can be bounded
from above by a quantity that is independent of both $m$ and $n$.
All the elements of $V^{-1/2}$ and of $-E$ are positive. Using the
techniques that led to  Eqs.~(\ref{v+})) we find
\begin{eqnarray*}
    \lvert V^{\pm 1/2}_{\vect{k} ,\vect{l} }\rvert &\ge &
    \frac{1}{2}\left(\frac{c}{2}\right)^{s(\vect{k} ,\vect{l})}
    \frac{1}{1-c^2}\, .
\end{eqnarray*}
As a consequence we have
\begin{equation*}
    \tr\left(-BE^\text{T}\right) \ge \sum_{\vect{k} ,\vect{l} }
    \left(\frac{1}{2}\left(\frac{c}{2}\right)^{s(\vect{k} ,\vect{l})}
    \frac{1}{1-c^2}\right)^2\, .
\end{equation*}
Now we employ counting methods analogous to those used in the
derivation for the upper bound we find an expression linear in the
area.  We take into account only contributions to the above sum
that correspond to the $2 d m^{d-1}$ oscillators that can be reached
in each step moving outwards orthogonal to the surface of the hypercube. We thus
obtain a lower bound proportional to the surface of the
hypercube $m^{\times d}$ for $m>m_{0}$ and appropriate $m_{0}$.
This concludes the proof.

\smallskip
In the following we will briefly describe possible extensions of
the above results that can be obtained by similar techniques,
including more general interactions, thermal states and classical
correlations in classical systems.

{\it `Squared interactions'. ---} The basic intuition behind the
entanglement-area dependence becomes most transparent for the
specific class of interactions for which the potential matrices
$V$ is of the form $V=W^2$ with a circulant band-matrix $W$. In
that case the covariance matrix of the ground state  is given by
$\gamma = W^{-1}\oplus W$. In  this case one arrives at
Eq.~\eqref{claim} since one can show that (i) the number of terms
contributing to the symplectic spectrum of the reduced covariance
matrix is linear in the number of degrees of freedom at the
boundary of the region, and (ii) the respective symplectic
eigenvalues are bounded from above and below independently of $n$
and $m$. Note that property (i) is equivalent to the existence of
a `disentangling' symplectic unitary transformation local to
inside and outside of the regions such that only oscillators near
to the boundary  remain entangled. Taking e.g.\
$V_1=\circM(1+2c^2,-2c,c^2,0,\dots,0,c^2,-2c)$ -- the case of
nearest-neighbor and  smaller next-to-nearest-neighbor
interactions -- allows to show that only the oscillators exactly
at the boundary contribute to the logarithmic negativity and that
$\lambda_1 (Q)\leq  2/(1-2c) -1$, with $Q$ being defined as in
Eq.~\eqref{q}. For the same interaction in $d=1$ spatial dimension
one can even exactly calculate the symplectic spectrum of the
reduced covariance matrix by means of a simple recursion relation.
In the limit $m\rightarrow \infty$ this results in  the two
non-vanishing symplectic eigenvalues $\mu_1=\mu_2 =
(1-c^2/q^2)^{-1/2}$, where $q=c+1/2 \pm (c+1/4)^{1/2}$.

{\it Entanglement and area in classical systems. ---} It should be
noted that, perhaps surprisingly, an `area-dependence' can also be
established analytically for classical correlations in classical
harmonic lattice systems \cite{Long}. It is noteworthy that this
result on classical  systems can be established most
economically using quantum techniques namely, mapping the problem
onto that of a quantum harmonic lattice with a squared interaction
as has been described above.

{\it Entanglement and area at finite temperature. ---} The
 property of squared interactions leading to effective disentanglement extends to
thermal states and therefore permits the proof of the linear
entanglement--area dependence for finite temperatures. In that case
operational entanglement measures such as the distillable
entanglement  have to be used. They can be bounded
from below by the hashing inequality and above again by the logarithmic
negativity \cite{Long}.

{\it Summary and outlook. ---} For certain harmonic lattice
Hamiltonians, e.g.\ discrete versions of the real Klein-Gordon
field, we have proven analytically that the degree of entanglement
between a hypercube and its environment can be bounded from above
and below by expressions proportional to the number of degrees of
freedom on the surface of the hypercube . This establishes rigorously a connection between
entanglement and area in this system. Intuitively, this originates
from the fact that one can approximately decouple the oscillators
in the interior and the exterior up to a band of the width of the
order of the correlation length of the system, which can be, as
outlined for the case of next-to-nearest neighbor coupling, equal
to just one lattice unit.

Our results can be extended to a wide variety of harmonic lattice
Hamiltonians, both quantum and classical, and a future publication
\cite{Long} will present details for more general interactions,
both ground and thermal states and a careful discussion of the
continuum limit, where the effective interaction strength is
modified.  These results in particular rely in an
essential way on the insights and techniques that have been
obtained in recent years in the development of a quantitative
theory of entanglement in quantum information science.

{\it Acknowledgements. ---} We warmly thank K.~Audenaert for input at earlier stages of this project, and J.~Oppenheim, T.~Rudolph and R.~F.~Werner for discussions. We would also like to thank J.~I.
Latorre and G.~Vidal for bringing the related Ref.\
\cite{Calabrese C 04} to our attention. This work was supported by
the DFG (SPP 1078), the EU (IST-2001-38877), the EPSRC (QIP-IRC),
and a Royal Society Leverhulme Trust Senior Research Fellowship.


\begin{thebibliography}{99}
%
\bibitem{Bombelli KLS 86}
L.\ Bombelli, R.\ K. Koul, J.\ Lee, and R.\ D. Sorkin, Phys.\
Rev.\ D {\bf 34}, 373 (1986).
%
\bibitem{Srednicki 93}
M.\ Srednicki, Phys.\ Rev.\ Lett.\ {\bf 71}, 666 (1993).
%
\bibitem{BlackHoleStuff}
J.\ M.\ Bardeen, B.\ Carter and S.\ W.\ Hawking, Commun.\ Math.\ Phys.\ {\bf 31}, 161 (1973);
J.\ D.\ Bekenstein, Lett.\ Nuovo Cimento\ {\bf 4}, 737 (1972);
G.\ 't~Hooft, Nucl.\ Phys.\ B\ {\bf 256}, 727 (1985); 
J.\ D.\ Bekenstein, Contemp.\ Phys.\ {\bf 45}, 31 (2004);
D.\ Kabat and M.\ J.\ Strassler, Phys.\ Lett.\ B\ {\bf 329}, 46 (1994); 
T.\ M.\ Fiola,  J.\ Preskill,  A.\ Strominger, and S.\ P.\
Trivedi, Phys.\ Rev.\ D {\bf 50}, 3987 (1994); G.\ Gour and A.\
E.\ Mayo, \textit{ibid.}\ {\bf 63}, 064005 (2001).
 C.\ Callan and F.\ Wilczek, Phys.\ Lett.\ B\ {\bf 333}, 55 (1994).
%
\bibitem{Holzhey 95}
C.\ Holzhey, F.\ Larsen and F.\ Wilczek, Nucl. Phys.\ B {\bf 424}, 443 (1995).
%
\bibitem{Casini 04}
H.\ Casini, Class.\ Quant.\ Grav.\ {\bf 21}, 2351 (2004). 
%
\bibitem{Yurtsever 03}
U.\ Yurtsever, Phys.\ Rev.\ Lett.\ {\bf 91}, 041302 (2003). 
%
\bibitem{Stelmachovic B 01}
P.\ Stelmachovic and V.\ Buzek, presented at a quantum 
information conference in Gdansk (July 2001);
P.\ Stelmachovic and  V.\ Buzek, Phys.\ Rev.\ A {\bf 70}, 032313 (2004).
%
\bibitem{Summers W 85}
S.\ J.\ Summers and R.\ F.\ Werner, Phys.\ Lett.\ A {\bf 110}, 257
(1985); H.\ Halvorson and R.\ Clifton, J.\ Math.\ Phys.\ {\bf 41},
1711 (2000); R.\ Verch and R.\ F.\ Werner, quant-ph/0403098;
B.\ Reznik, A.\ Retzker, and J.\ Silman, quant-ph/0310058; B. Reznik,
A.\ Retzker and J.\ Silman, J.\ Mod.\ Opt.\ {\bf 51}, 833 (2004).
%
\bibitem{Audenaert EPW 02}
K.\ Audenaert, J.\ Eisert, M.\ B.\ Plenio,\ and R.\ F.\ Werner,
Phys.\ Rev.\ A {\bf 66}, 042327 (2002).
%
\bibitem{Wolf VC 04}
M.\ M.\ Wolf,  F.\ Verstraete,  and 
J.\ I.\ Cirac, Phys.\ Rev.\ Lett.\ {\bf 92}, 087903 (2004).
%
\bibitem{Botero R 04}
A.\ Botero and B.\ Reznik, Phys. Rev. A {\bf 70}, 052329 (2004).
%
\bibitem{Fannes HM 03}
M. Fannes, B. Haegeman, and M. Mosonyi, J. Math. Phys. {\bf 44}, 6005 (2003).
%
\bibitem{Latorre} 
G.\ Vidal, J.\ I.\ Latorre, E.\ Rico, and A.\ Kitaev, Phys.\ Rev.\ Lett.\ {\bf 90} 227902 (2003); 
J.\ I.\ Latorre, E.\ Rico, and G.\ Vidal, Quant.\ Inf.\ Comp.\
{\bf 4}, 48 (2004);
J.\ I.\ Latorre, C.\ A.\ Lutken, G.\ Vidal, quant-ph/0404120. 
%
\bibitem{Eisert P 03}
J.\ Eisert and M.\ B.\ Plenio, Int.\ J.\ Quant.\ Inf.\ {\bf 1}, 479 (2003).
%
\bibitem{Neg}
J.\ Eisert and M.\ B.\ Plenio, J.\ Mod.\ Opt.\ {\bf 46}, 145 (1999);
J.\ Eisert (PhD thesis, Potsdam, February 2001);
G.\ Vidal and R.\ F.\ Werner, Phys.\ Rev.\ A {\bf 65}, 032314 (2002);
K.\ Audenaert, M.\ B.\ Plenio, and J.\ Eisert,
Phys.\ Rev.\ Lett.\ {\bf 90}, 027901 (2003).
%
\bibitem{Matrix}
R.\ A.\ Horn and C.\ R.\ Johnson, {\it Matrix Analysis} (Cambridge
University Press, Cambridge, 1985).
%
\bibitem{Long}
Forthcoming publication by the same authors in different order.
%
\bibitem{Calabrese C 04}
P.\ Calabrese and J.\ Cardy, J.\ Stat.\ Mech.\ Th.\ Exp P00406 (2004). 

\end{thebibliography}
\end{document}